\documentclass[a4paper]{jpconf}
\usepackage{graphics}
\usepackage{psfrag}
\newcommand{\imsc}
{\address{$^2$Institute of Mathematical Sciences, CIT Campus
Taramani, Chennai 600 113, India.}}

\begin{document}
  
 \title
{Opinion dynamics model with domain size dependent dynamics: 
novel features and new universality class}

\author
 {Soham Biswas$^1$, Parongama Sen$^1$, Purusattam Ray$^2$}
\address{$^1$Department of Physics, University of Calcutta,
92 Acharya Prafulla Chandra Road, Kolkata 700009, India}


\imsc
\ead{soham.physics@gmail.com, parongama@gmail.com, ray@imsc.res.in}
\begin{abstract}

A model for opinion dynamics (Model I) has been recently introduced  in which 
the binary opinions of the individuals are determined according to the size
of their neighboring domains (population having the same opinion).
The coarsening dynamics of the equivalent Ising model shows power law behavior 
and has been found to belong to a new universality class with the dynamic exponent $z=1.0 \pm 0.01$
and persistence exponent $\theta \simeq 0.235$ in one dimension. The critical behavior has been 
found to be robust for a large variety of annealed disorder that has been studied. 
Further, by mapping Model I to a system of random walkers in one dimension 
with a tendency to walk towards their nearest neighbour with probability  $\epsilon$, we 
find that for any $\epsilon > 0.5$, the Model I dynamical behaviour is prevalent at long times.


\end{abstract}


\section{Introduction}
\label{sec:Introduction}

Sociophysics has emerged as one of the important areas of research during recent times. The 
concepts of statistical physics find application to many situations that occur in a social system  
with the assumption that individual free will or feelings do not take crucial role in these situations
\cite{bkcbook,Castellano}. One of the major issues that has attracted lots of attention
is how  opinions evolve in a social system. Starting from random initial opinions, dynamics 
often  leads to a consensus which means a major fraction of the population support a certain 
cause, for example a motion or a candidate in an election etc.

Simulating human behaviour by models effectively implies quantifying the outcome of the behavior 
by suitable variables having continuous or discrete values. Different dynamical rules are 
proposed for the evolution of these variables, depending on how these variables change with time 
following social interactions. Thus, a social system can be treated like a physical system. For example, 
in case of opinion dynamics, if the opinions have only discrete binary values, the 
social system can be regarded as a magnetic system of Ising spins. 

In this context, Schelling model \cite{Schell}, proposed  in 1971, seems to be the very first model 
of opinion dynamics. Since then, a number of models describing the formation of
opinions in a social system have been proposed \cite{opinion}. 
While on one hand these models attempt an understanding of how a society 
behaves and social viewpoints evolve, on the other hand, these provide rich complex
dynamical physical systems suitable for theoretical studies. 

Dynamics of complex systems has become a subject of extensive research from several aspects. 
For many such systems, e.g., traffic or agent based models, one cannot define a  conventional
Hamiltonian or energy function. The only method by which one can study the 
steady state behaviour of such systems is by looking at the long time dynamics. 
Nonequilibrium dynamics involves the evolution of a system from a completely random initial 
configuration and associated with this evolution are   several phenomena of interest like domain growth
or persistence that have been studied, for example, in spin systems. Since in many sociophysics 
model, one can have variables analogous to  spin variables, these phenomena can be readily studied 
here. An important objective  is  to identify dynamical universality classes by estimating the relevant 
dynamical exponents. 
 
Another point of interest in studying dynamics is that  many systems may have identical equilibrium behaviour 
but behave differently as far as  dynamics is concerned. For example, Ising spin dynamics with or without 
conservation belong to different dynamic universality class although their equilibrium behaviour is identical.

Apart from the dynamical behaviour, different kinds of phase transitions have also been observed
in these models by introducing suitable parameters. One such phase transition can be from a 
homogeneous society where everyone has the same opinion to a heterogeneous  one with mixed 
opinions \cite{phase-tr}.

Change in the opinion of an individual takes place in different ways in different models. For example
in the Voter model \cite{vote}, an individual simply follows the opinion of a randomly chosen neighbour 
while in the Sznajd model \cite{Sznajd}, the opinion of one or more individuals are changed following more 
complicated rules.

In this article, we review the dynamical studies in a recently introduced model \cite{sbps1}
(to be referred  to as Model I henceforth) and its variants in which a new rule of updating (discussed 
in detail in section 2) is introduced. In one dimension, Model I can be visualized as an Ising spin chain 
(the binary opinions are represented by Ising spins) with a Glauber-like dynamics, 
where a spin only at the domain boundary can  flip. In this new model, the state of the spin is determined 
by the state of the neighboring domain larger in size (detailed description is given in the next section). 
Model I shows strikingly different dynamical behavior compared to 
known models of opinion dynamics or spin dynamics. Several observables show power law decay and 
the exponents strongly suggest a new universality class. 

The introduction of disorder in various forms have also been 
considered which shows that the Model I dynamical behaviour is not affected by annealed disorder. 
We also report (in section 3) new results when a mapping of this model is made to an equivalent reaction diffusion  model
where the walkers move towards their nearest neighbour with probability $\epsilon$. 
Once again Model I dynamical behaviour is seen to exist at long times for any $\epsilon > 0.5$ (which means a bias 
towards the nearest neighbour) showing the 
extreme robustness of the model. 


\section{Description of the Model I}

In  a model of opinion dynamics, the key feature is
the interaction of the individuals. Usually, in all the models, it is assumed that
an individual
is influenced by its nearest neighbours.
Model I  is a one dimensional 
model of binary opinion in which the dynamics 
is dependent on the {\it {size}} of the neighbouring domains as well.
Here an individual changes his/her opinion in two situations: 
first when the two  
neighbouring domains have opposite polarity,  and in this case 
the individual simply follows the opinion of
the neighbouring domain with the  larger size.
This case may arise only when the individual is at the boundary of the two 
domains.
An individual also 
changes his/her opinion when both the  neighbouring domains have an opinion 
which opposes his/her original opinion, i.e., the individual is  
sandwiched between two domains of same polarity.
It may be noted that for  the second case, 
the size of the neighbouring domains is irrelevant.
When  the two neighbouring domains are of the same size but have opposite polarity,
the individual will change his/her orientation with fifty percent probability.

\begin{figure}[ht]
\begin{center}
{\resizebox*{6.3cm}{!}{\includegraphics{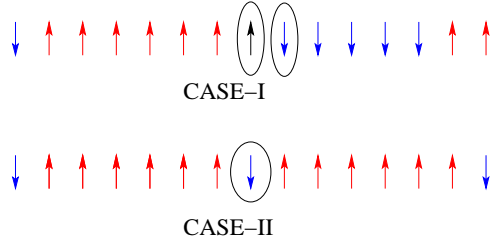}}}
\caption{Dynamical rules for Model I: in both cases the encircled spins may change state; in case I, 
the boundary spins 
will follow the opinion of the left domain of up spins which  will grow. 
 For case II,  the 
down spin between the two up spins will flip irrespective of the size of the neigbouring domains.}
\label{model}
\end{center}
\end{figure}

The binary opinions can  be represented by a system of Ising spins where the up and down states
correspond to the two possible opinions. 
The  two  rules followed in the dynamical evolution  
  in the equivalent spin model 
are  shown schematically 
in Fig. \ref{model} as case I and  II. 
In the first case the spins at the boundary   
between two domains will choose the state of the 
left side domain (as it is  larger in size). For the 
second case the down spin  flanked by  two neighbouring  up spins 
will flip.

 The main idea in Model I is that the size of a domain represents
 a quantity analogous to `social pressure' which is expected 
  to be proportional to the  number of  people 
supporting a cause. An individual, sitting at the domain boundary, is most exposed to
the competition between  opposing pressures and  gives in to the larger one.
This is what happens in case I shown in Fig.\ref{model}. 
The interaction in case II on the other hand implies that it is difficult to 
stick to   one's opinion if the entire (immediate) neighbourhood opposes it.  

Defining the dynamics in this way, one immediately notices that case II corresponds to what would
happen for  spins in a nearest neighbour Ferromagnetic Ising Model (FIM) in which the dynamics
at zero temperature is simply an energy minimization scheme.
However,   the boundary spin in the FIM
behaves differently in case I; it may or may not flip as the energy remains same.
 In the present model, the dynamics is
deterministic even for the boundary spins
(barring the rare 
instance when the two neighbourhoods have the same size in which case 
the spin flips with fifty percent probability).

In this
model, the important condition of changing one's opinion is the size of the
neighbouring domains which is not fixed  either in time or space.
This is the unique feature of this model. 
In the most familiar  
models of opinion dynamics like the Sznajd model \cite{Sznajd} or the voter model \cite{vote},
one takes the effect of nearest neighbours within a given radius and
even in the case of models defined on networks \cite{v-network}, 
the influencing neighbours may be
nonlocal but always fixed in identity.

In the equivalent spin model, if $L_{+}$ is the number of up spins 
and $L_{-}$ is the number down spins, 
the order parameter is defined as $m=\vert L_{+}-L_{-} \vert/L $.
This is identical to the (absolute value of) magnetization.
Starting from a random initial configuration, the dynamics in Model I showed that it 
leads to a final state with $m=1$, i.e. a homogeneous state where all spins have 
the same value (either +1 or -1). It is not difficult to understand this result; in absence 
of any fluctuation, the dominating neighbourhood (domain) 
simply grows in size ultimately spanning the entire system.

\begin{figure}[ht]
\begin{center}
{\resizebox*{8cm}{!}{\includegraphics{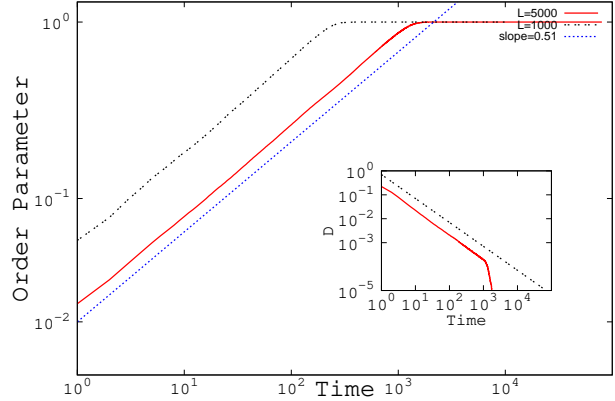}}}
\caption{Variation of the  order parameter $m$ with time for two different system sizes along with
a straight line (slope 0.51) shown in a log-log plot.
Inset shows the decay of fraction of domain wall $D$ with time.}
\label{NP_Dw}
\end{center}
\end{figure}

In the spin picture, the dynamics can be described in terms of the movement of the domain 
walls and as the dynamics progresses, number of domain walls goes on decreasing.
Monte Carlo simulations showed that the domain dynamics and the dynamics of the order 
parameter obey conventional power law  variations:  the fraction of domain walls
$D(t) \propto  t^{-1/z}$ with $z = 1.00 \pm 0.01 $ and  order parameter $m(t) \propto t^{1/2z}$
(Fig. \ref{NP_Dw}).

\begin{figure}[ht]
\begin{center}
{\resizebox*{8cm}{!}{\includegraphics{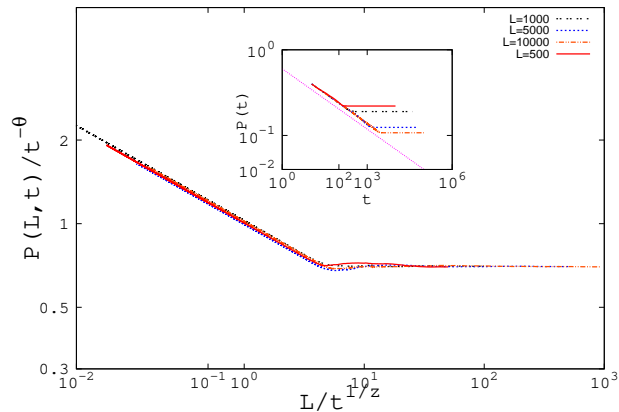}}}
\caption{The collapse of scaled persistence probability versus scaled time using $\theta=0.235$ and $z=1.04$ is shown for
different system sizes.  Inset shows 
the unscaled data.}
\label{per_noparam}
\end{center}
\end{figure}

The persistence measure (i.e., probability that a spin has not  
flipped till time $t$) \cite{derrida1}
showed the familiar power law behaviour:
$P(t) \propto t^{-\theta}$, 
where $\theta$ is the  persistence exponent.
For finite system of size $L$, $P(t,L)$ is known to  behave as \cite{puru,biswas_sen} 
\begin{equation}
P(t,L) \propto t^{- \theta}f(L/t^{1/z}),
\label{fss}
\end{equation}
and at large times,  the persistence probability saturates at a value $ \propto L^{-\alpha}$. 
Therefore,  for
 $x <<1$, $f(x)  \propto x^{-\alpha}$ with $\alpha = z\theta$.  For large $x$,
$f(x)$ is a constant. Estimate for both  $z$ and $\theta$ using the above
scaling form showed that 
 $\theta =0.235\pm0.003$, 
and  a $z$ value $1.04 \pm 0.01$ (Fig.  \ref{per_noparam}).

Model I therefore showed a novel dynamical behaviour with 
values of $z$ and $\theta$   quite different from those of  
the one dimensional Ising model \cite{Derrida} and other 
opinion/voter dynamics models \cite{stauffer2,sanchez,shukla}. 
Specifically in the Ising model,  
$z=2$ and $\theta = 0.375$   
and for the   Sznajd model
the persistence exponent is  equal to that of the Ising model.

\subsection{Model I with a cutoff}

As a variation of the  Model I, one can  introduce a  cutoff \cite{sbps2} denoted by a  
parameter $p$,  such that the maximum size of the neighbouring 
domains an individual  can sense is given by $R = pL/2$ in a one dimensional system of $L$  agents
with periodic boundary condition. $p=1$ corresponds to the original Model I.
A finite cutoff (i.e., $p < 1$) puts a restriction on the domain sizes which  may  correspond 
to geographical, political, cultural boundaries etc. The case with uniform cutoff signifies 
that all the individuals have same kind of restriction. The case with random cutoffs, which is 
perhaps closer to reality, has also been studied briefly. Once again one 
can represent the system by Ising spins.  

Whenever $R$ is kept finite, that is, $R$ does not scale with system sizes (implying $p \rightarrow 0$ in the 
thermodynamic limit), the dynamics leads to the equilibrium
configuration of all spins up/down and the
dynamic exponents also turn out to be identical to those
corresponding to the nearest neighbour Ising values (i.e.,
$\theta = 0.375 $ and $ z = 2$).

\begin{figure}[ht]
\begin{center}
{\resizebox*{16cm}{!}{\includegraphics{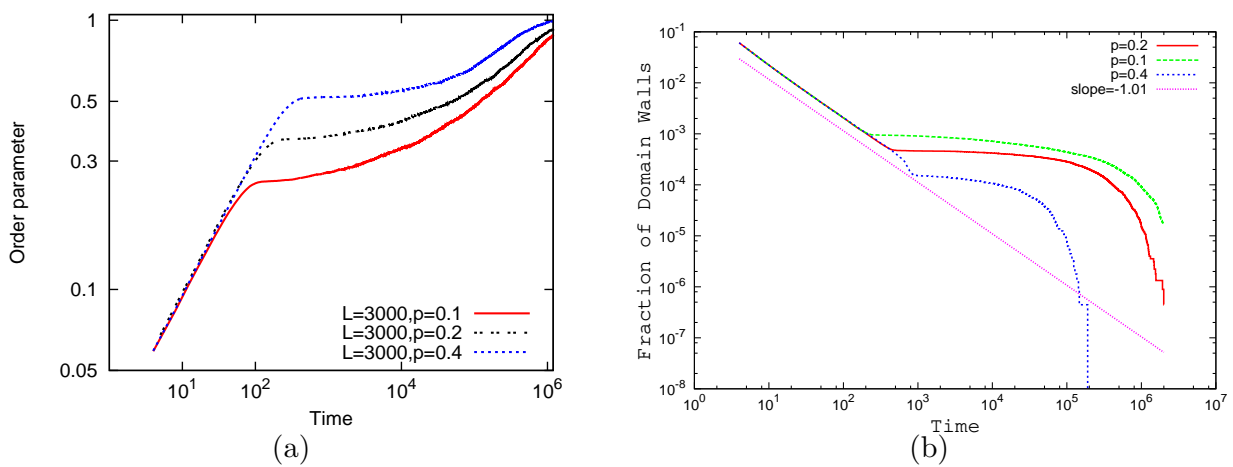}}}
\caption{Variation of the (a) order parameter and the (b) fraction of domain walls with time for different finite cutoff 
$p=0.1,0.2,0.4$ ($L=3000$).  }
\label{mag_domain}
\end{center}
\end{figure}

The results with a nonzero $p$ showed that there is a crossover phenomena in dynamics.
For an initial time $t_1 = pL/2$, the dynamics remains Model I-like as the cutoff 
does not affect the dynamics till this time. Beyond this time, 
the dynamics becomes diffusive. Such a crossover is, however, not apparent from the 
log-log plots of the relevant dynamical quantities (Fig.  \ref{mag_domain} ). This is because the crossover occurs between
two different types of phenomena. The first one is pure coarsening in which domain walls
prefer to move towards their nearest neighbours as in Model I and one gets the expected
power-law behaviour. There is a correlation length which is growing; as long as this length
is smaller than $R$ one will observe ballistic growth (as in Model I). When
the correlation length is larger than $R$ one observes domain diffusion. The
diffusive behavior is not evident in the coarsening process because ballistic
dynamics ``leaves'' the system into a non-typical configuration which is evidently far from those on diffusion paths.
At $t_1$, some special configurations are generated in which only a few 
domain walls remain, the density
of domain walls going to zero in the thermodynamic limit.
Therefore beyond $t_1$, the second phenomenon involves pure diffusion of these domain walls which remain non-interacting up
to large times. Only when the distance between two domain walls become $\leq R$ the Model I like dynamics prevails
again before the two are annihilated. This takes place over a much shorter time scale such that the physical quantities
attain saturation values ($m=1$ and $D=0$) very fast in the last part of the dynamics (Fig.  \ref{mag_domain} ).

The only dynamic exponent in the diffusive regime is the diffusion
exponent $z = 2$, which is distinct from the growth exponent $z = 1$. So the two dynamic
exponents not only differ in magnitude, but they are also connected to distinct phenomena.
Using both numerical data
and some analytical arguments, it could be shown that the behaviour at $t>t_1$
is indeed diffusive. The saturation time to reach equilibrium was found to be
\begin{equation}
t_{sat} = a pL + b (1-p)^3 L^2,
\label{tsat}
\end{equation}
which also showed that for $L \to \infty$, $z=2$.  

The persistence behaviour, however, does not show any power law behaviour corresponding 
to the diffusive behaviour, i.e., it does not decay like   
$t ^ {-0.375}$  as in a  usual reaction diffusion system. This is because of the special configurations in which the 
system is left at time $t_1$. However, the long time behaviour showed that the persistence probability saturates in time and decays 
with the system size as $L^{-\alpha}$ with  $\alpha = 0.235$, which is the Model I value. In fact one can obtain a 
collapse for the persistence data using $\alpha =0.235$ in both the regions $ t<t_1$ (with $z=1$) and $t>t_1$ (with $z=2$) 
for any $p\neq 0$.
Thus the crossover phenomenon occurs with a novel characteristic behaviour.

\subsection{Effect of disorder}

The Model I described so far has no disorder, which can be introduced in several 
possible ways. We discuss three cases in the following.

\subsubsection{Effect of rigidity parameter:}

Since every individual is not expected to 
succumb to social pressure,   
   Model I can be  modified by  introducing a parameter $\rho$ called rigidity 
coefficient which denotes the probability that people are completely rigid 
and never change their opinions \cite{sbps1}. This means there are  
$\rho N$ rigid individuals  (chosen randomly at time $t=0$),
  who retain their initial state 
throughout the time evolution. 
Thus the disorder is quenched in nature. Such rigid individuals had been
 considered earlier in \cite{galam}.

The limit $\rho=1$
corresponds to the unrealistic  noninteracting case when no time evolution 
takes place; 
$\rho=1$ is in fact a trivial fixed point.
For other values of $\rho$, the system evolves to a equilibrium state.

The time evolution changes drastically in nature (compared to Model I) with the 
introduction of $\rho$. 
All the dynamical variables like
order parameter,  fraction of domain wall and persistence
attain a saturation value at a rate which increases with 
$\rho$.  Power law variation with time can only be observed for $\rho < 0.01$ 
with the exponent values same as those for $\rho = 0$.
The saturation or equilibrium values on the other hand show
the following behaviour:
\begin{eqnarray}
&&m_s\propto N^{-\alpha_1}\rho^{-\beta_1}\nonumber\\
&&D_s \propto \rho^{-\beta_2}\nonumber\\
&&P_s  =  a + b\rho^{-\beta_3}  \label{}
\end{eqnarray}
where in the last equation $a$ is a constant  $\simeq 0.06$ independent of $\rho$.
The values of the exponents are $\alpha_1 = 0.500 \pm 0.002,$ 
$\beta_1 = 0.513 \pm 0.010$, $\beta_2= 0.96 \pm 0.01$ and  $\beta_3 = 0.430 \pm 0.01$.

%
It can be naively assumed that the $N\rho$ rigid individuals will dominantly appear at the 
domain boundaries such that  in the first order approximation (for a fixed 
population),
 $D \propto 1/\rho$. This would give $m \propto 1/\sqrt{\rho}$ 
indicating  $\beta_1 = 0.5$ and $\beta_2 =1$.   
The numerically obtained values are in fact quite close
to these estimates.


The results obtained above can be explained in the following way:
with $\rho \neq 0$, 
the domains cannot grow freely  and domains with both
kinds of opinions survive making the equilibrium 
$m_s$ less than unity.
Thus  the society becomes heterogeneous for any $\rho > 0$ when people
do not follow the same opinion any longer. 
The variation of $m_s$ with $N$ shows  
that $m_s \rightarrow 0$ in the thermodynamic limit for $\rho > 0$. 
 Thus not only does the society become  heterogeneous at  the onset
of $\rho$, it goes to a completely disordered state 
analogous to the
paramagnetic state in magnetic systems.  
Thus one may conclude that a phase transition from a ordered state with $m=1$
to a disordered state ($m=0$) takes place for $\rho = 0^+$.

%


Since the role of $\rho$ is similar to domain wall pinning, one can
introduce a depinning probability factor $\mu$ which in this system  
represents the probability for  rigid individuals  to become non-rigid
during each Monte Carlo step. $\mu$  relaxes the rigidity
criterion in an annealed manner in the sense 
that the identity of the individuals who become non-rigid is not fixed (in time). If  
$\mu=1$, one gets back  Model I whatever
be the value of $\rho$, and therefore $\mu=1$ signifies a line of (Model I) 
fixed points,
where the dynamics leads the system to a homogeneous state.

\begin{figure}[ht]
\begin{center}
{\resizebox*{7cm}{!}{\includegraphics{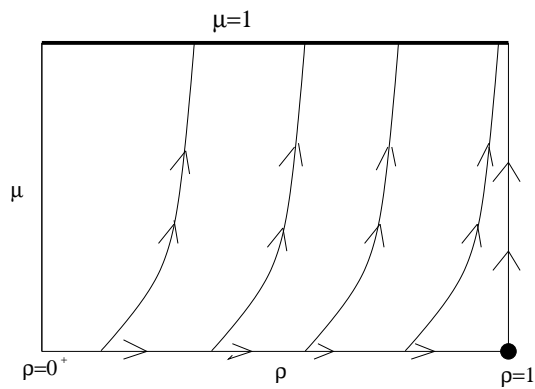}}}
\caption{
The flow lines in the $\rho-\mu$ plane:  
Any non-zero value of $\rho$ with $\mu=0$ drives
the system  to the disordered fixed point $\rho=1$. 
Any nonzero value of $\mu$ drives it to the 
ordered state ($\mu=1$, which is a line of fixed points) 
for all values of $\rho$.}
\label{flow}
\end{center}
\end{figure}

With the introduction of $\mu$, one has effectively  a lesser fraction  $\rho^\prime$  of
rigid people in the society,  where
\begin{equation}
\rho^\prime = \rho(1-\mu).
\end{equation}
The difference  from the previous model with quenched rigidity is, of course, that 
this effective fraction of   rigid individuals 
is not fixed in identity (over time). Thus when $\rho \neq 0, \mu \neq 0$, we have a system
in which there are both quenched and annealed disorder.
It is observed that for any  nonzero value of $\mu$, 
the system once again evolves to a homogeneous state ($m=1$) for 
all values of $\rho$. Moreover, the dynamic behaviour is also same as Model I 
with the exponent $z$ and $\theta$  having identical 
values.
This shows that the nature of randomness is crucial as 
 one cannot simply 
replace a system with parameters \{$\rho \neq 0, \mu \neq 0$\} 
by one with only quenched randomness
 \{$\rho^\prime \neq 0$, $\mu^\prime=0$\} as in the latter   case
one would end up with a heterogeneous society.
We therefore conclude that the annealed disorder
wins over the quenched disorder;
  $\mu$ effectively drives the system to
 the  $\mu=1$ fixed point  for 
any value of $\rho$. This is shown   schematically in  a flow diagram (Fig. \ref{flow}).
It is worth remarking that it looks very similar to the flow
diagram of the one dimensional Ising model with nearest neighbour 
interactions in a longitudinal field and finite temperature.

\subsubsection{Effect of thermal-noise like disorder : }

In thermodynamical systems, the effect of thermal disorder is 
a highly important issue. In Model I which describes a social system, 
a disorder analogous to thermal noise has been  introduced \cite{ps}.

Let $d_{up}$ and $d_{down}$ be the sizes of the two neighbouring
domains of type up and down of a spin at the domain boundary (excluding itself). In Model I, probability $P(up)$ that the said spin is up  
is  1 if $d_{up} > d_{down}$, 0.5 if  $d_{up} = d_{down}$ and zero otherwise.
In the simplest possible way to introduce disorder, one may 
take the  probability of a boundary spin to be up as  
$P(up) = d_{up}/ (d_{up}+ d_{down})$.
However, there is no parameter
controlling the stochasticity here and moreover,  the 
results are identical to the original Model I and therefore this
kind of stochasticity is not of much importance. 

In order to introduce a noise like parameter which can be tuned,  
the probability that a spin at the domain boundary is up  was taken to be 
\begin{equation}
P(up) \propto e^{\beta(d_{up}- d_{down})},
\end{equation}
and it is down with probability 
\begin{equation}
P(down) \propto e^{\beta(d_{down}- d_{up})}.
\end{equation}
The normalized probabilities are  therefore 
$P(up) =  \exp{\beta\Delta}/(\exp(\beta\Delta) + \exp(-\beta\Delta))$ and  
 $P(down)=1-P(up)$,
where $\Delta = (d_{up}- d_{down}) $.

\begin{figure} [ht]
\begin{center}
\rotatebox{270}{\resizebox*{4.7cm}{!}{\includegraphics{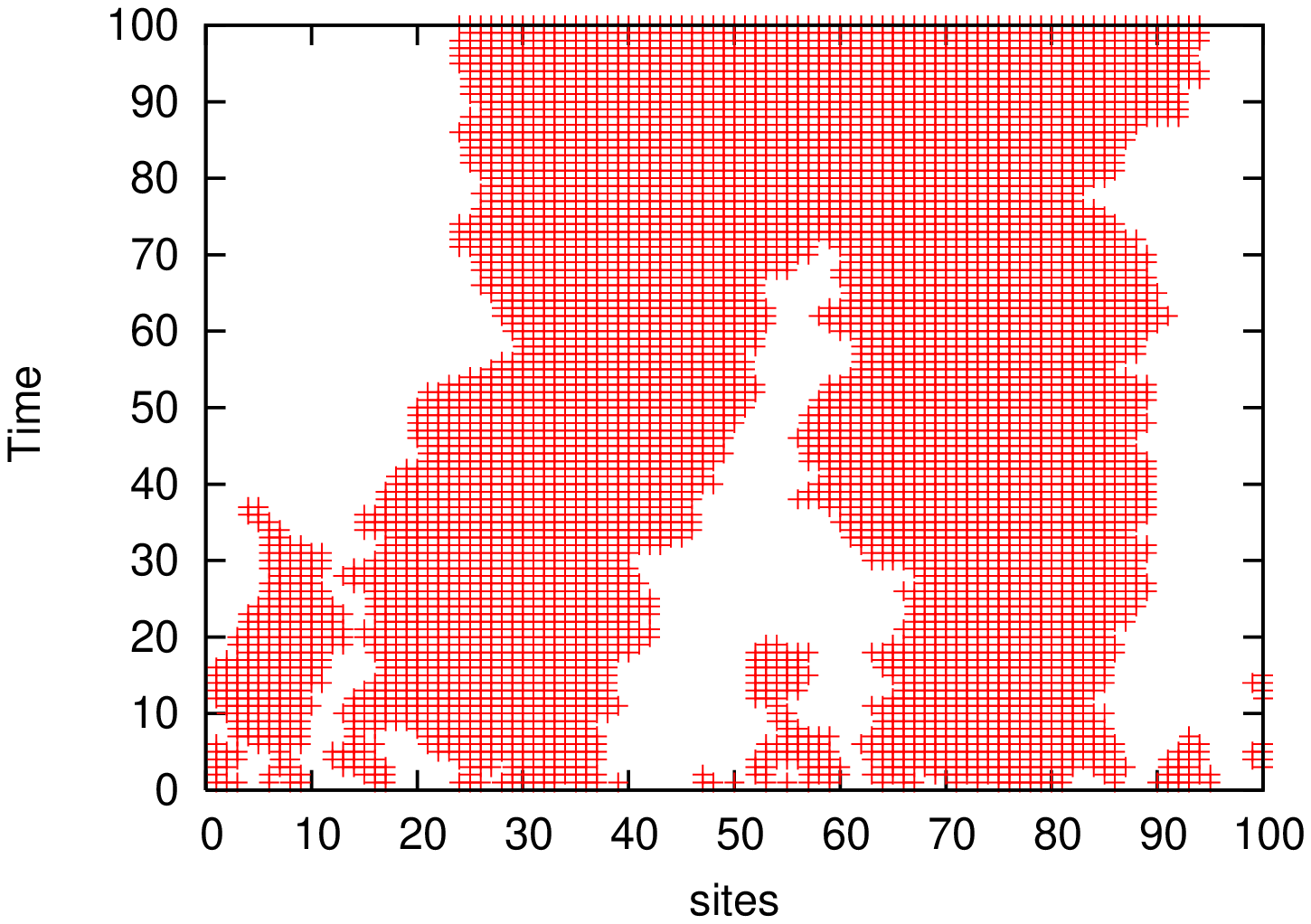}}}
\rotatebox{270}{\resizebox*{4.7cm}{!}{\includegraphics{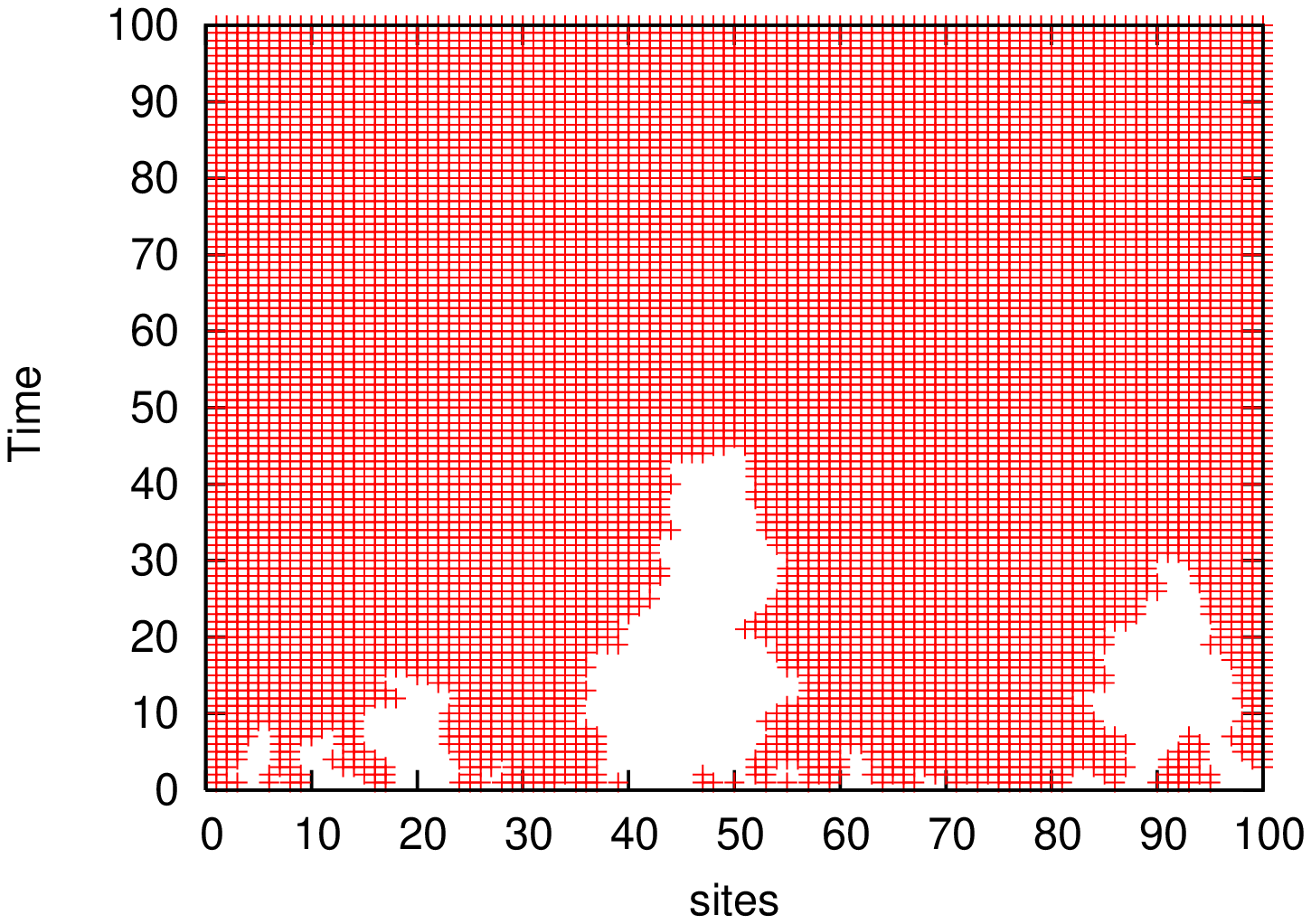}}}

\rotatebox{270}{\resizebox*{4.7cm}{!}{\includegraphics{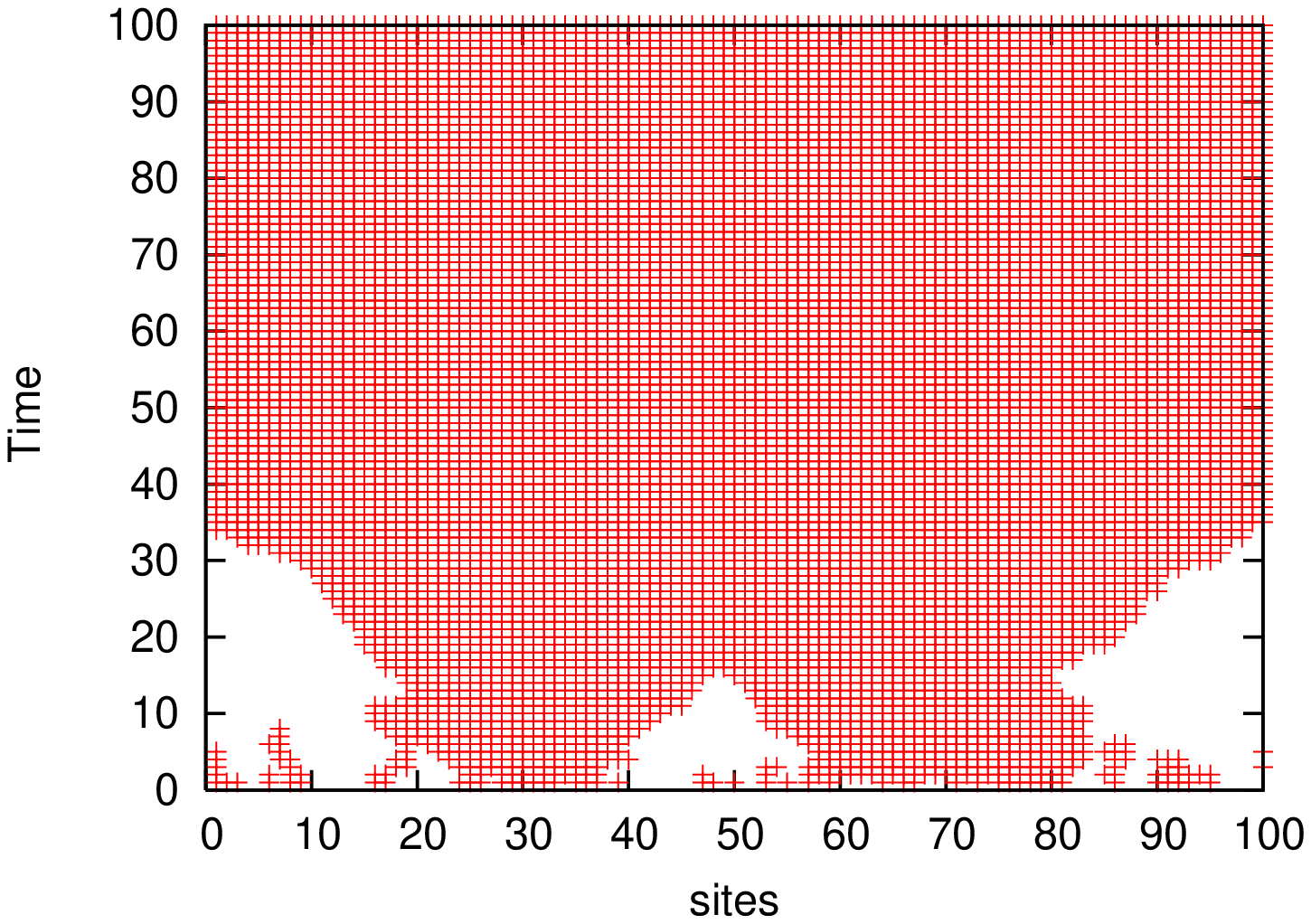}}}
\rotatebox{270}{\resizebox*{4.7cm}{!}{\includegraphics{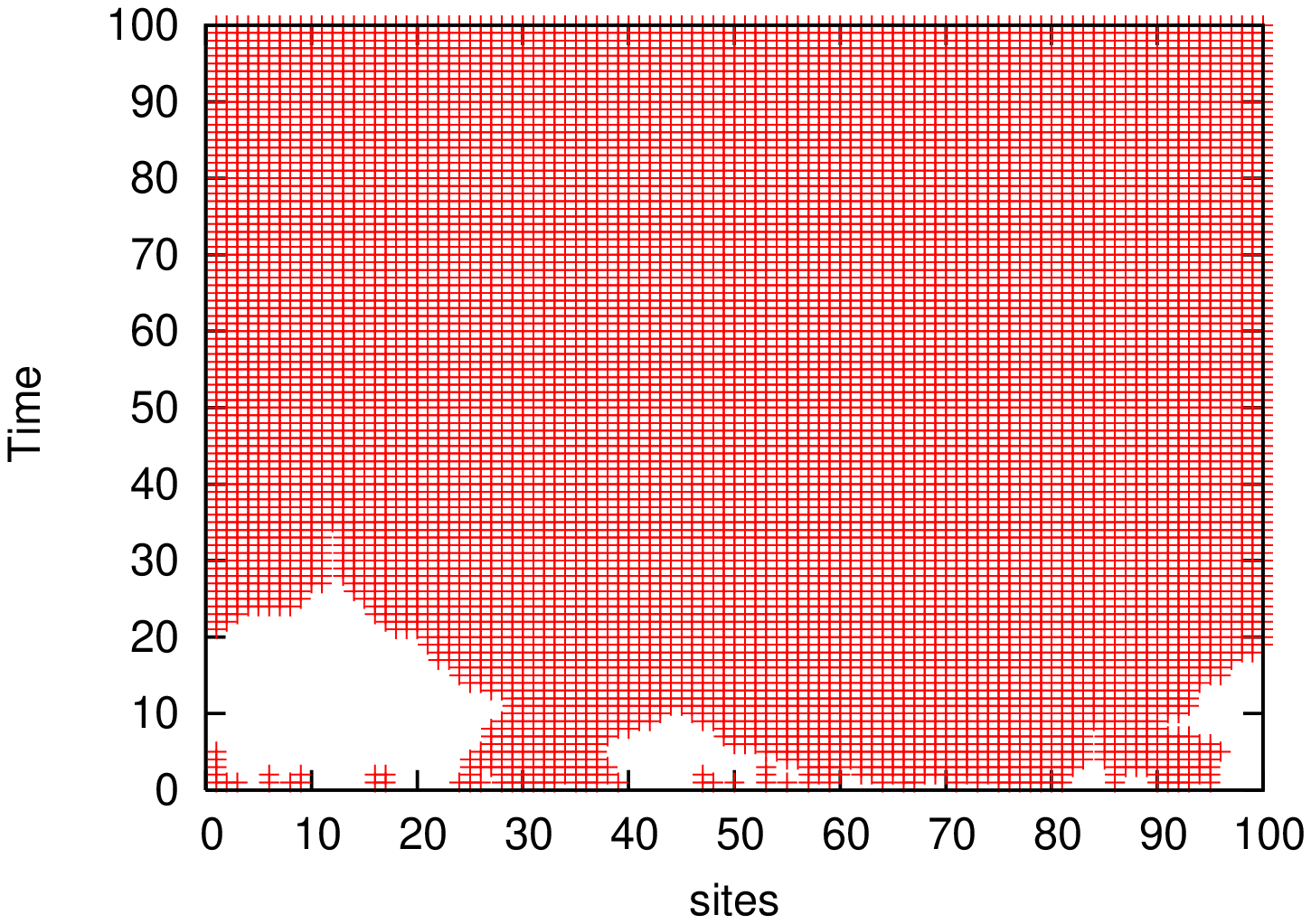}}}
\caption{Snapshots of the system in time for different values of $\beta = 0.0, 0.005, 0.1$, and $\beta \to \infty$ (top to bottom) 
 showing that even for very small  non-zero $\beta$, the system equilibrates
towards a homogeneous state much faster compared to the $\beta = 0 $ case. These snapshots are for a 
$L=100$ system.}
\label{dissnap}
\end{center}
\end{figure}

Obviously, $\beta \to \infty$ corresponds to  Model I while 
letting $\beta =0$ we have  equal probabilities of the up and down states,  making it equivalent to the zero temperature dynamics of the 
nearest neighbour Ising model. Since the equilibrium states for the extreme values $\beta \to \infty$ and 
$\beta =0$  are homogeneous  
 (all up or all down states), it is expected that for all values of $\beta$ they will remain so  as is indeed the case.

It is useful to show the snapshots of the evolution of the system over
time for different $\beta$ (Fig.\ref{dissnap}): to be noted is the fact that 
for any non-zero $\beta$ however small, the system equilibrates very fast compared to the Ising limit $\beta =0$. 

The detailed dynamical studies in fact showed that the system belongs to the Model I 
dynamical class  
 for any finite $\beta$ and a dynamic transition takes place exactly at $\beta = 0$.
Certain subtleties in this model need to be mentioned:
with respect to Model I, $\beta=0$ is the maximum noise and its inverse may be thought of an effective temperature. 
On the other hand, from the Ising model viewpoint, $\beta$ plays the role of noise.  
However it is not equivalent to thermal fluctuations which can affect the state of any spin. 
With $\beta$, flipping of spins  can still occur at the domain boundaries only. 
Hence, even with this noise, the equilibrium behaviour 
is not disturbed for any value of $\beta$ (even for $\beta \to \infty$ which corresponds to Model I) 
while in contrast, any non-zero temperature can destroy the order of a one dimensional Ising model. 

\subsubsection{Model with random cutoffs : }

Previously we have discussed the case when a uniform cutoff
 is introduced to the system. 
One can introduce randomness in the cutoff parameter $p$ varying from 0 to 1,
(chosen randomly from a uniform distribution)
and associated with each individual a different value of $p$ \cite{sbps2}.
 The randomness is quenched as the value of
$p$ assumed by any individual  is fixed for all times. 
In this case,  
 the variation of the magnetization, domain walls and persistence show
power law scalings with exponents  corresponding to  Model I only for an initial range of time; 
however, calculation of  $z$ from the saturation times gives $z=1$. 
The later time behaviour appears to deviate from the Model I behaviour
and further studies are required to analyses the exact dynamical behaviour.

\begin{figure} [ht]
\begin{center}
\rotatebox{0}{\resizebox*{15cm}{!}{\includegraphics{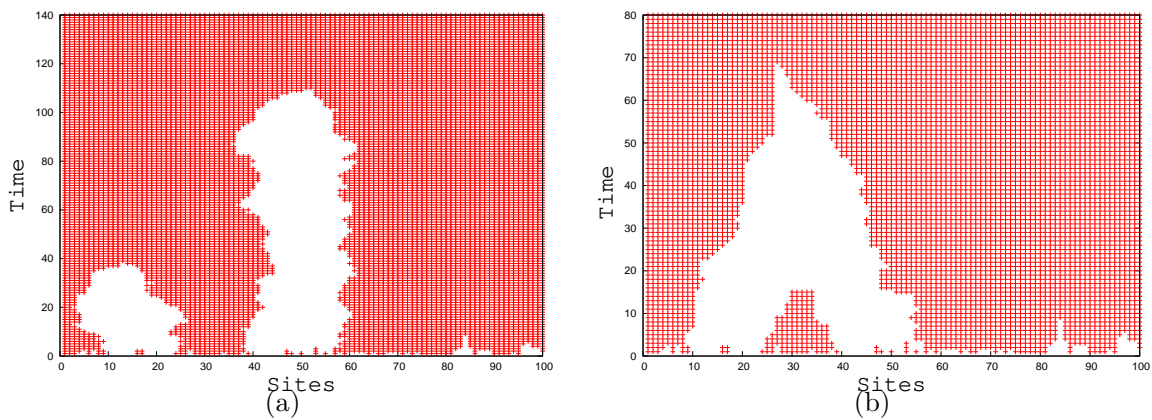}}}
\caption{Snapshots of the system in time for (a) random cutoff parameter $p$ varying from 0 to 1, and 
(b)  uniform cutoff parameter $p$.
  These snapshots are for a 
$L=100$ system.}
\label{cutsnap}
\end{center}
\end{figure}

Even snapshots of the system (when compared to the uniform cutoff 
or infinite cutoff cases) do not help in understanding the phenomenon much (Fig \ref{cutsnap}).

\section{Mapping of the opinion dynamics model to reaction diffusion  system}

The opinion dynamics model discussed so far are models where the dynamics is described 
in terms of the Ising spins that mimic the binary opinions an individual can have. In this model, 
a spin deep inside a domain does not flip. The dynamics is governed by the flipping of the spins only 
at the domain walls. The dynamics, in this respect, is reminiscent of the zero temperature Glauber 
dynamics of the kinetic Ising model. The motions of  the domain walls can be viewed as the 
motions of the particles $A$ with the reaction $A + A \rightarrow \emptyset$.  This means the 
particles are walkers and when two particles come on top of each other they are annihilated.
The annihilation reaction ensures domain coalescence and coarsening.  
Unlike that in Glauber Ising model, the walkers $A$ corresponding to Model I do not perform 
random walks. These walkers move {\it ballistically towards their nearest neighbours}. 
This bias, as we have seen before, gives rise to a new universality class than that of 
conventional reaction diffusion system \cite{privman}.

We have also studied $A + A \rightarrow \emptyset$ model with the particles $A$ performing
random walk with a bias $\epsilon$ towards their nearest neighbors. 
We have taken $\epsilon$ as the probability that  a walker walks 
towards its nearest neighbour. Clearly, $\epsilon  = 0.5$ corresponds 
to usual reaction diffusion system with the particles performing random walk. 
On the other hand, $\epsilon=1$ is equivalent to our Model I as has been described above. 
We have studied the dynamics of the reaction diffusion system for different values of $\epsilon$ 
in the range $[0^{+},1.0]$. 

\begin{figure} [ht]
\begin{center}
\rotatebox{270}{\resizebox*{6.5 cm}{!}{\includegraphics{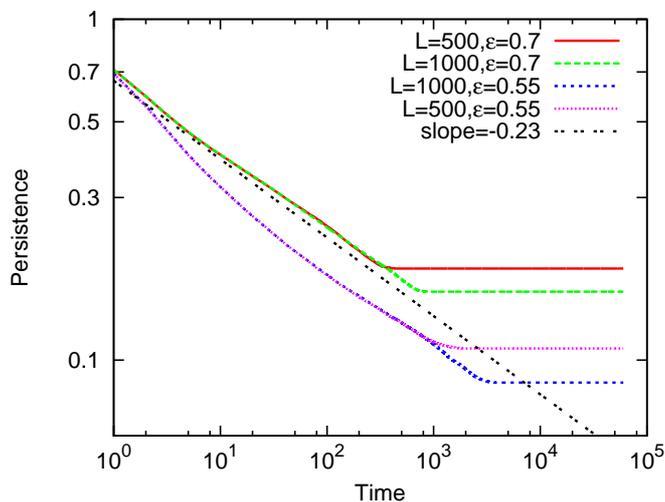}}}
\caption{Decay of persistence  with time for $\epsilon =0.7$ and $\epsilon =0.55$}
\label{pereplt5}
\end{center}
\end{figure}

\begin{figure} [ht]
\begin{center}
\rotatebox{270}{\resizebox*{6.5cm}{!}{\includegraphics{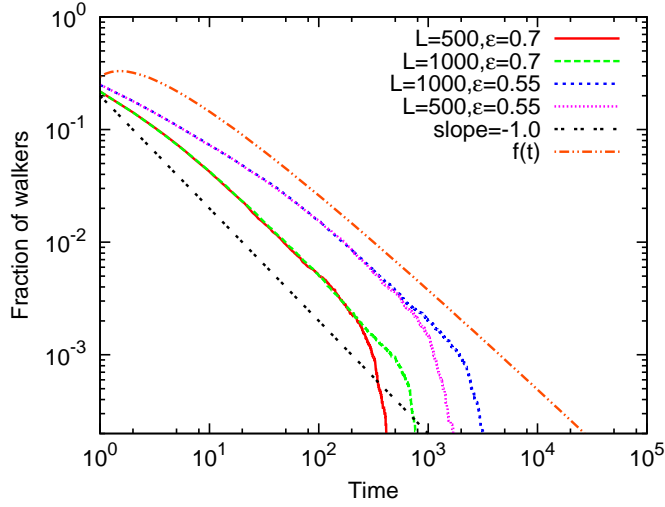}}}
\caption{Decay of number of walkers with time for $\epsilon =0.7$ and $\epsilon =0.55$. There is a logarithmic 
correction to scaling for the value of $\epsilon =0.55$. The form of $f(t)$ is $t^{-1}(1+\alpha \log(t))$ with 
$\alpha = 3.92$}
\label{wlkeplt5}
\end{center}
\end{figure}

In reaction diffusion systems, the growth of domains is given by the 
number of surviving walkers. Persistence $P(t)$ in these systems is defined as the fraction of sites 
unvisited by any of the walkers $A$ till time $t$. Figures \ref{pereplt5} and \ref{wlkeplt5} show the decay of 
persistence and fraction of walkers with time for different values of $\epsilon > 0.5$. 

We find that for $\epsilon > 0.5$, the Model I behaviour is observed, namely: $z \simeq 1.0$ and 
$\theta \simeq 0.235$,  with some possible correction to the  scaling which becomes weaker as 
$\epsilon$ is increased.  For example, there is a 
logarithmic correction to scaling for the decay of the fraction of domain 
walls which takes the form $t^{-1}(1+\alpha (\epsilon) \log(t))$ where 
 $\alpha (\epsilon) \to 0$ as   $\epsilon \to 1$. 
One can compare the above model with the cases discussed in section 2.2.3, where
the introduction of stochastic dynamics also occurred with 
a bias towards the larger domain. In case of thermal disorder, the parameter 
comparable to $\epsilon$ is $\beta$. In the present case, the
exact sizes of the domains do not matter (which is important for the case with $\beta$) but the results are consistent
in the sense that any bias towards the larger domain (or nearest walker) makes the system behave 
like Model I. 

\begin{figure} [ht]
\begin{center}
\rotatebox{270}{\resizebox*{6.5cm}{!}{\includegraphics{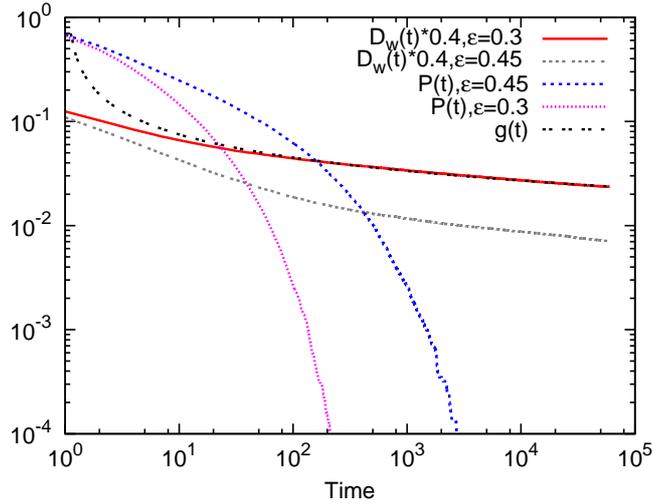}}}
\caption{Decay of persistence and number of walkers with time for $\epsilon =0.3$ and $\epsilon =0.45$. 
The form of $g(t)$ is $ a(1/ \log(t))$ where $a$ is any constant.}
\label{eplt5}
\end{center}
\end{figure}

In this model, we have  also studied the $\epsilon < 0.5$ region 
where the opposite happens, the walker has a bias towards the further neighbour.
Obviously domain annihilations take place very slowly now, even slower than 
$1/\log(t)$ and the dynamics continues for very long times. Consequently, 
the persistence probability no longer shows a power law variation now  but falls 
exponentially to zero (Fig \ref{eplt5}).

\section{Summary and Discussions}

We have reviewed the dynamics of a recently proposed opinion dynamics model (Model I and 
its variations) \cite{sbps1}. Model I  in one dimension belongs to a new dynamical universality 
class with novel dynamical features not encountered in any previous models of dynamic spin system
or opinion dynamics. In the corresponding reaction diffusion system $A + A \rightarrow \emptyset$, we have 
introduced a probability $\epsilon$ of random walkers $A$ moving towards their nearest
neighbors. $\epsilon = 0.5$ corresponds to the particles $A$ performing unbiased random walks and the 
system belongs to the dynamical universality class of zero temperature Glauber Ising model. We find that 
for $\epsilon > 0.5$, the system still shows power law behavior of domain growth and persistence but with 
a universality class of that of Model I. For $\epsilon < 0.5$,  the domain grows logarithmically and the 
persistence decays exponentially in time. 

We have discussed quite a few variations on Model I with quenched and annealed disorders
and cutoffs in the interaction range. In presence of quenched disorders, like the presence of few 
rigid spins which do not flip, the dynamics changes drastically.  The dynamics ends up in 
heterogeneous equilibrium phases, which is fully disordered in the sense that no consensus
can be reached as the order parameter goes to zero in the
thermodynamic limit, with identical behaviour irrespective of the percentage of the 
rigid spins in the system. However, the dynamics shows extreme robustness against annealed 
disorder. We have also discussed the crossover phenomena in the dynamics in presence of cutoffs 
in the range over which a spin recognizes its neighboring domains.

All the  models discussed in the present article can easily be extended to higher dimensions and its 
universality class determined. Phase transitions occurring at 
non-extreme values of suitably defined parameters may also be expected in higher dimensions.\\

Acknowledgements:
SB and PS acknowledge financial support from DST
project SR/S2/CMP-56/2007. Partial
computational help has been provided by UPE project. 
SB acknowledges the hospitality of IMSc where part of the work was done.
PR acknowledges the Department of Physics, University of Calcutta for hospitality.

\end{document}